\documentclass[12pt,a4paper,twoside]{article}
\usepackage{amsmath,amsfonts,amstext,amsthm,amssymb}   
\usepackage{natbib}
\usepackage{graphics}
\bibpunct[,]{(}{)}{;}{a}{,}{,}   

\def\eqref#1{(\ref{#1})}
\theoremstyle{definition}
\newtheorem{prop}{Proposition}
\newtheorem{exam}{Example}
\newcommand{\with}{\enspace\text{with}\enspace}
\newcommand{\half}{\frac{1}{2}}
\newcommand{\R}{\mathbb R}
\newcommand{\gegen}{\longrightarrow}
\renewcommand{\phi}{\varphi}
\newcommand{\bm}[1]{\mbox{\boldmath$#1$}}
\newcommand{\vx}{\bm{v}}
\newcommand{\vhat}{\bm{\hat{\vx}}}
\newcommand{\xs}{\ensuremath{(x_1,\ldots,x_n)}}
\newcommand{\argminv}{\mathop{\text{argmin}}_{\vx}}
\newcommand{\rmax}{r_\text{max}}
\newcommand{\rcut}{c}
\newcommand{\IF}{\mathrm{IF}}
\newcommand{\dr}{\,\mathrm{d}r}
\newcommand{\rp}{\rho^*_{\mathrm{eff}}}
\newsavebox{\mycirc}
\sbox{\mycirc}{\scalebox{1.5}{$\circ$}}
\newsavebox{\myxmark}
\sbox{\myxmark}{\scalebox{0.9}{$\times$}}
\newcommand{\includescaled}[2]{\scalebox{#1}{\includegraphics{#2}}}
\newcommand{\Tend}{T_\mathrm{end}}
\newcommand{\weg}[1]{}

\begin{document}

\begin{center}
\Large\bf 
Redescending M-estimators and Deterministic\\ Annealing, with Applications to Robust Regression\\and Tail Index Estimation
\end{center}

\vspace*{0.1cm}

\begin{center}
\large 
Rudolf~Fr\"uhwirth and Wolfgang~Waltenberger\\
Institute of High Energy Physics,
\\Austrian Academy of Sciences, Vienna, Austria
\end{center}

\vspace*{0.1cm}

\begin{quote}
{\bf Abstract:} {A new type of redescending M-estimators is constructed, based on data augmentation with an unspecified outlier model. Necessary and sufficient conditions for the convergence of the resulting estimators to the Huber-type skipped mean are derived. By introducing a temperature parameter the concept of deterministic annealing can be applied, making the estimator insensitive to the starting point of the iteration. The properties of the annealing M-estimator as a function of the temperature are explored. Finally, two applications are presented. The first one is the robust estimation of interaction vertices in experimental particle physics, including outlier detection. The second one is the estimation of the tail index of a distribution from a sample using robust regression diagnostics.}

{\bf Zusammenfassung:}
{Ein neuer Typ von wiederabsteigenden M-Sch\"atzern wird konstruiert, ausgehend von Datenerweiterung mit einem unspezifizierten Ausrei\ss{}ermodell. Notwendige und hinreichende Bedingungen f\"ur die Konvergenz zu Hubers
``Skipped-mean''-Sch\"atzer werden angegeben. Durch Ein\-f\"uhrung einer Temperatur kann die Methode des ``Deterministic Annealing'' angewendet werden. Der Sch\"atzer wird dadurch unempfindlich gegen die Wahl des Anfangspunkts der Iteration. Die Eigenschaften des Sch\"atzers als Funktion der Temperatur werden untersucht. Schlie\ss{}lich werden zwei Anwendungen vorgestellt. Die erste ist die robuste Sch\"atzung von Wechselwirkungspunkten in der experimentellen Teilchenphysik, einschlie\ss{}lich der Erkennung von Ausrei\ss{}ern. Die zweite ist die Sch\"atzung des ``Tail index'' einer Verteilung aus einer Stichprobe mittels robuster Regressionsdiagnostik.}

{\bf Keywords:} Redescending M-estimator, Deterministic annealing, Robust regression, Regression diagnostics, Tail index estimation.
\end{quote}

\section{Introduction}

M-estimators were first introduced by~\citet{Huber} as robust estimators of location and scale. Their study in terms of the influence function was undertaken by Hampel and co-workers~\citep{Hampel}. Redescending M-estimators are a special class of M-estimators. They are widely used for robust regression and regression clustering, see e.g.~\citet{Muller2004} and the references therein. According to the definition in~\citet{Hampel}, the $\psi$-function of a redescending M-estimators has to disappear outside a certain central interval. Here, we merely demand that the $\psi$-function tends to zero for $|x|\gegen\infty$. If $\psi$ tends to zero sufficiently fast, observations lying farther away than a certain bound are effectively discarded. Redescending M-estimators are thus particularly resistant to extreme outliers, but their computation is afflicted with the problem of local minima and a resulting dependence on the starting point of the iteration.

The problem of convergence to a local minimum can be cured by combining the iterative computation of the M-estimate with a global optimization technique, namely deterministic annealing. For a review of deterministic annealing and its applications to clustering, classification, regression and related problems see~\citet{Rose1998} and the references therein. To the best of our knowledge, the combination of M-estimators with deterministic annealing has been proposed only by~\citet{Li1996}. It will be shown below, however, that his annealing M-estimators have infinite asymptotic variance at low temperature, a feature that we deem to be undesirable in certain applications.

The purpose of this note is to construct a new type of redescending M-estimators with annealing that converge to the Huber-type skipped mean~\citep{Hampel} if the temperature $T$ approaches zero. The starting point is a mixture model of data and outliers. Data augmentation is used to formulate an EM algorithm for the estimation of the unknown location of the data. The EM algorithm is then interpreted as a redescending M-estimator that can be combined with deterministic annealing in a natural way (Subsections~\ref{sec:construction} and~\ref{sec:temperature}). The most important case is a normal model for the data, but other models are possible. In Subsection~\ref{sec:nonnormal} conditions are derived under which the corresponding M-estimator converges to the skipped mean. Section~\ref{sec:ntype}
explores the properties of the annealing M-estimator with a normal data model and illustrates the effect of deterministic annealing on a simple example with synthetic data. Section~\ref{sec:applications} presents two applications of the annealing M-estimator: first, robust regression applied to the problem of estimating an interaction vertex in experimental particle physics; and second, regression diagnostics applied to the problem of estimating the tail index of a distribution from a sample.

\section{Redescending M-estimators with\\ Deterministic Annealing}

This section shows how a new type of redescending M-estimators can be constructed via data augmentation. The estimators are then generalized by introducing a temperature parameter so that deterministic annealing can be applied. The case of data models other than the normal one is discussed, and conditions for convergence to the skipped mean are derived.

\subsection{Construction via Data Augmentation}\label{sec:construction}

The starting point is a simple mixture model of data with outliers, with the p.d.f. 
\begin{gather}
h(x)=p\cdot f(x;\mu,\sigma) + (1-p)\cdot g(x).\label{eq:model}
\end{gather}
By assumption, $f(x;\mu,\sigma)$ is the p.d.f.\ of the normal distribution with location $\mu$ and scale $\sigma$ and can be written as
\begin{gather*}
f(x;\mu,\sigma)=\phi(r),\quad\text{with}\quad r=(x-\mu)/\sigma,
\end{gather*}
$\phi(.)$ being the standard normal density. The distribution of the outliers, characterized by the density $g(x)$, is left unspecified.

Now let \xs{} be a sample of size $n$ from the model in Eq.~\eqref{eq:model}. The sample is augmented by a set of indicator variables $I_j, j=1,\ldots,n$, where $I_j=0\,(1)$ indicates that $x_j$ is an inlier (outlier).
If the scale $\sigma$ is known, the location can be estimated by the EM algorithm~\citep{emalgo}. In this particular case, the EM algorithm is an iterated re-weighted least-squares estimator, the weight of the observation $x_j$ being equal to the posterior probability that it is an inlier. The latter is given by Bayes' theorem:
\begin{gather}
P(I_j=0|x_j)=\frac{P(x_j|I_j=0)\cdot P(I_j=0)}{P(x_j|I_j=0)\cdot P(I_j=0)+P(x_j|I_j=1)\cdot P(I_j=1)}
\label{eq:post}
\end{gather}
As we do not wish to specify the outlier distribution, we resort to a worst case scenario and set $P(I_j=0) = P(I_j=1) =0.5$. In addition we require that in the vicinity of $\mu$, an observation should be an inlier rather than an outlier, so $P(I_j=1|x)\leq P(I_j=0|x)$ for $|x-\mu|/\sigma\leq c, c>0$, where $c$ is a cutoff parameter. This can be achieved by setting the prior probability $P(x_j|I_j=1)$ that $x_j$ is an outlier to $P(\mu+c\sigma|I_j=0)=\phi(c)$. The posterior probability $P(I_j=0|x_j)$ then reads:
\begin{gather}
P(I_j=0|x_j)=\frac{f(x_j;\mu,\sigma)}{f(x_j;\mu,\sigma)+\phi(c)}
=\frac{\phi(r_j)}{\phi(r_j)+\phi(c)},
\label{eq:post2}
\end{gather}
where $r_j=(x_j-\mu)/\sigma$. If $r_j=c$, the posterior probabilities of $x_j$ being an inlier or an outlier, respectively, are the same. 

If the inlier density is normal the EM algorithm is tantamount to an iterated weighted mean of the observations:
\begin{align*}
\mu^{(k+1)}&=\sum_{j=1}^n w_j^{(k)} x_j /\sum_{j=1}^n w_j^{(k)},\quad\text{with}\\
  w_j^{(k)}&=\frac{\phi(r_j^{(k)})}{\phi(r_j^{(k)})+\phi(c)},\quad\text{and}\\
  r_j^{(k)}&=(x_j-\mu^{(k)})/\sigma
\end{align*}
The iterated weighted mean can also be interpreted as an M-estimator of location~\citep{Huber}, with 
\begin{gather*}
\psi(r;c)=\frac{r\phi(r)}{\phi(r)+\phi(c)},\quad\text{and}\quad \rho(r;c)=\int \psi(r;c)\dr.
\end{gather*}
This interpretation allows us to analyze the estimator in terms of its influence function and associated concepts such as gross-error sensitivity and rejection point.

\subsection{Introducing a temperature}\label{sec:temperature}

\begin{figure}[htb]
\begin{center}
{\includegraphics{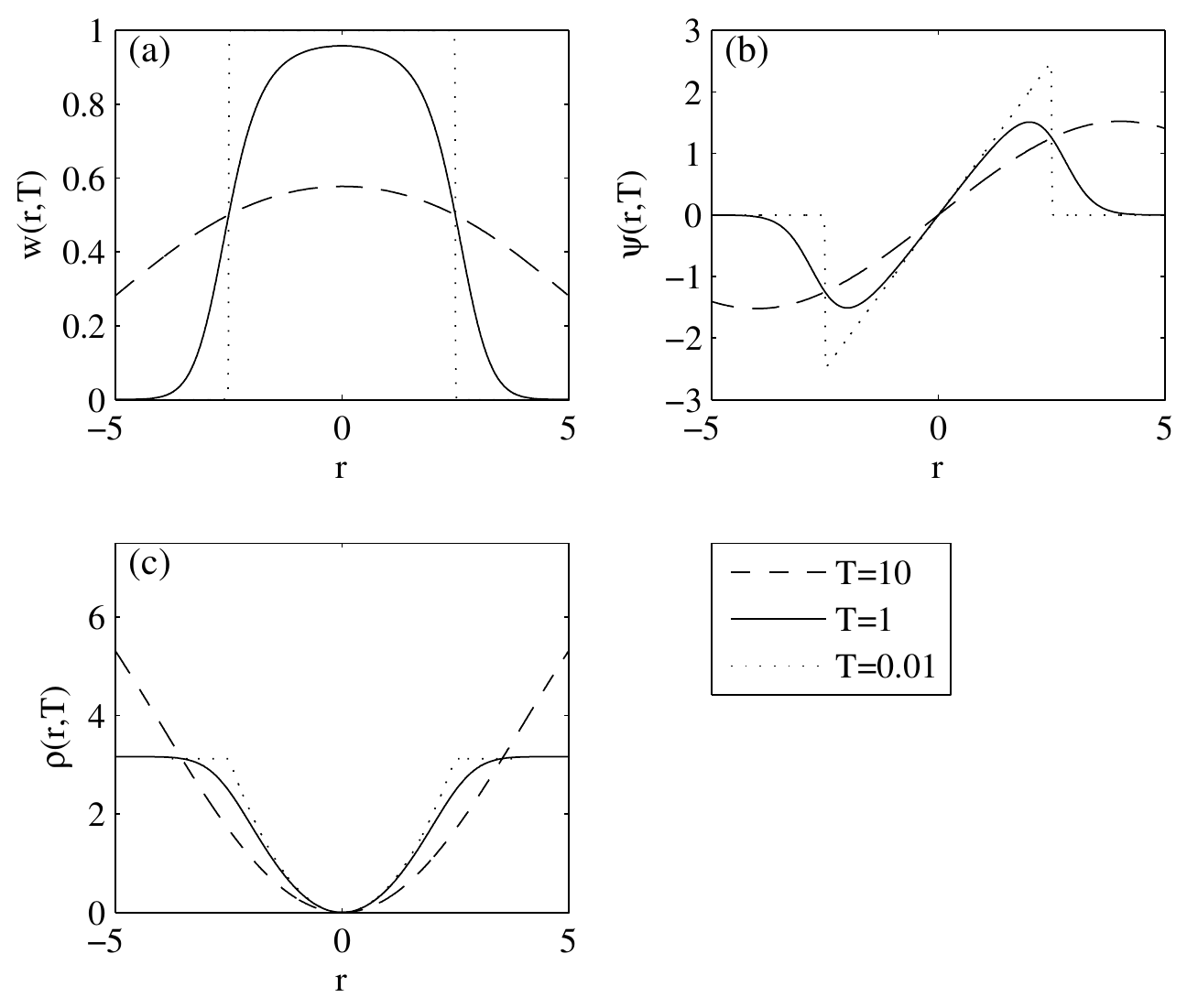}}
\caption{N-type M-estimator: (a) $w(r;c,T)$, (b) $\psi(r;c,T)$ and (c) $\rho(r;c,T)$ for $c=2.5$ and $T=10,1,0.01$.}\label{fig:N-type}
\end{center}
\end{figure}

The shape of the score function $\rho(r;c)$ can be modified by introducing a temperature parameter $T$ into the weights. This allows to improve global convergence by using the technique of deterministic annealing~\citep{Rose1998,Li1996}. The modified weights are defined by
\begin{gather}
w(r;c,T)=\frac{\phi(r/\sqrt{T})}{\phi(r/\sqrt{T})+\phi(c/\sqrt{T})}=\frac{\exp(-r^2/2T)}{\exp(-r^2/2T)+\exp(-c^2/2T)}.\label{eq:w}
\end{gather}
The redescending M-estimator with this weight function is called a normal-type or N-type M-estimator. Its $\psi$-function is given by 
\begin{gather*}
\psi(r;c,T)=\frac{r\,\exp(-r^2/2T)}{\exp(-r^2/2T)+\exp(-c^2/2T)},\label{eq:normal_psi}
\end{gather*}
and its $\rho$-function by
\begin{gather}
\rho(r;c,T)= \frac{r^2}{2}  - T \ln \left(
\exp({r^2}/{2T}) + \exp({c^2}/{2T}) \right)
+ T \ln \left( 1 + \exp({c^2}/{2T}) \right).\label{eq:normal_rho}
\end{gather}
Figure~\ref{fig:N-type} shows the weight function, the $\psi$-function and the $\rho$-function of the N-type M-estimator for three different temperatures $(T=10,1,0.01)$. Note that Eq.~\eqref{eq:normal_rho} is not suitable for the numerical computation of $\rho(r;c,T)$ if $T$ is very small. A numerically stable version of Eq.~\eqref{eq:normal_rho} is given by:
\begin{gather*}
\rho(r;c,T)= 
\begin{cases}
\displaystyle \frac{r^2}{2}+T\ln\frac{1+\exp(-c^2/{2T})}{1+\exp((r^2-c^2)/{2T})} & \text{if $|r|<c$}, \\[\bigskipamount]
\displaystyle \frac{c^2}{2}+T\ln\frac{1+\exp(-c^2/{2T})}{1+\exp((c^2-r^2)/{2T})} & \text{if $|r|>c$}.
\end{cases}
\end{gather*}

If the temperature increases, the weight drops more slowly as a function of $r$. In the limit of infinite temperature we have
\begin{gather*}
\lim_{T\gegen\infty}w(r;c,T)=\frac{1}{2},
\end{gather*}
for all $c$, and the M-estimator degenerates into a least-squares estimator. If the temperature drops to zero, the weight function converges to a step function.
\begin{prop}\label{prop:1}
Let $H(\cdot)$ be the unit step function (Heaviside function) with the additional convention $H(0)=1/2$. Then
\begin{gather*}
\lim_{T\gegen 0} w(r;c,T)= H(c-r).\tag*{$\Box$}
\end{gather*}
\end{prop}
\noindent Proposition~\ref{prop:1} follows from the more general Proposition~\ref{prop:2} below.

\subsection{Non-normal data models}\label{sec:nonnormal}

The density $\phi$ used in defining the weight function in Eq.~\eqref{eq:w} need not be the standard normal density. In fact, every unimodal continuous symmetric density $f(x)$ with location 0, scale 1 and infinite range generates a type of redescending M-estimators. The behaviour of the weight function $w_f(r;c,T)$ at low temperature is determined by the tail behaviour of $f(x)$, as described by the concept of regular variation at infinity~\citep{Seneta}. We recall that a function $f(x): [0,\infty)\gegen(0,\infty)$ is called regularly varying at infinity with index $\xi\in\R$ if it satisfies
\begin{gather*}
\lim_{x\gegen\infty}\frac{f(\lambda x)}{f(x)}=\lambda^{\xi}\quad\text{for any $\lambda>0$.}
\end{gather*}
If $f(x)$ is a probability density function, $\xi$ has to be in the interval $(-\infty,-1)$. The definition can be extended in the obvious sense to $\xi=-\infty$. If $\xi=-\infty$,
\begin{gather*}
\lim_{x\gegen\infty}\frac{f(\lambda x)}{f(x)}=
\begin{cases} 0&\text{for $\lambda>1$,}\\
              \infty&\text{for $\lambda<1$.}
\end{cases}
\end{gather*}
In this case $f(x)$ is also called rapidly varying at infinity~\citep{Seneta}. Normal densities are rapidly varying at infinity, as are all densities with exponential tails.

\begin{prop}\label{prop:2}
\mbox{\ }\\[-\baselineskip]
\begin{enumerate}
\item[(a)]
Let $H(\cdot)$ be as in Proposition~\ref{prop:1}. $f(r)$ is rapidly varying at infinity if and only if
\begin{gather*}
\lim_{T\gegen 0} w_f(r;c,T) = H(c-r).
\end{gather*}
for all $c>0$.
\item[(b)]
$f(r)$ is regularly varying at infinity with index $\xi\in\R$ if and only if
\begin{gather*}
\lim_{T\gegen 0} w_f(r;c,T) = \frac{r^\xi}{r^\xi+c^\xi}=\frac{c^{-\xi}}{r^{-\xi}+c^{-\xi}}.
\end{gather*}
for all $c>0$.\hfill{$\Box$}
\end{enumerate}
\end{prop}
\noindent The proof is omitted, but can be obtained from the authors on request.
\weg{
\begin{proof}
If we set $x=r/\sqrt{T}$ we have
\begin{gather*}
\lim_{T\gegen 0} w_f(r;c,T)=\left[1+\lim_{x\gegen\infty}\frac{f(x\cdot c/r)}{f(x)}\right]^{-1}.
\end{gather*}
\begin{enumerate}
\item[(a)]
If $f(r)$ is rapidly varying at infinity, the right hand side is equal to $H(c-r)$, and vice versa.
\item[(b)]
If $f(r)$ is regularly varying at infinity with index $\xi\in\R$, then
\begin{gather*}
\left[1+\lim_{x\gegen\infty}\frac{f(x\cdot c/r)}{f(x)}\right]^{-1}=\left[1+(c/r)^\xi\right]^{-1}=
\frac{r^\xi}{r^\xi+c^\xi}.
\end{gather*}
If the equality holds for all $c>0$, $f(r)$ is regularly varying at infinity with index $\xi$.\qedhere
\end{enumerate}
\end{proof}
}

\begin{figure}
\begin{center}
{\includegraphics{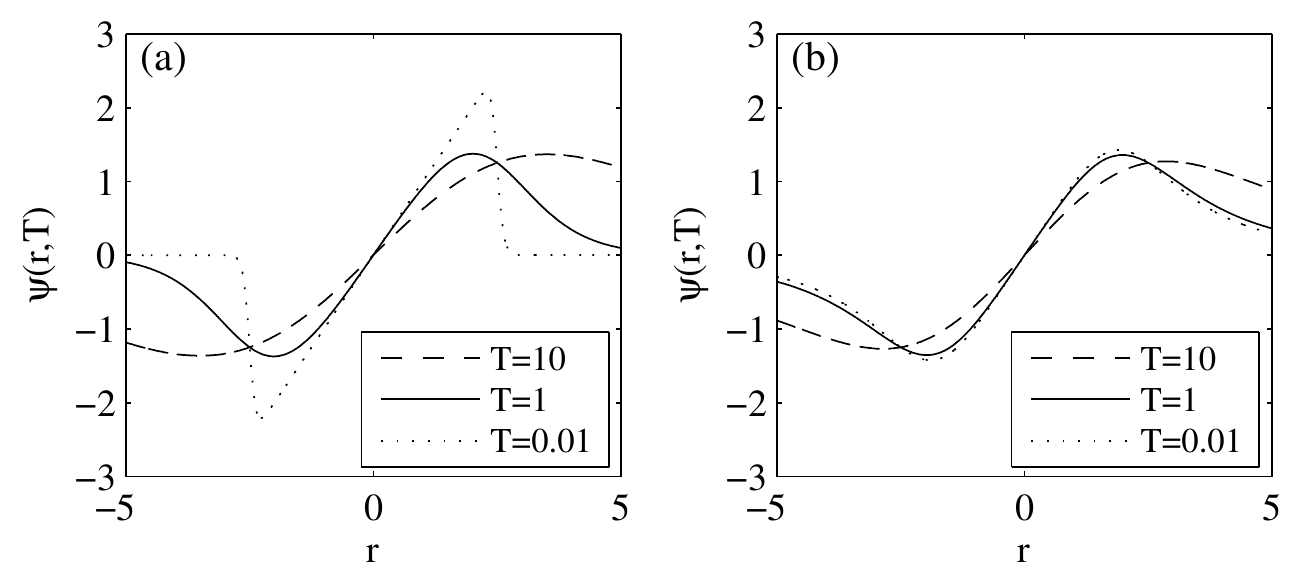}}
\caption{$\psi(r;c,T)$ of redescending M-estimators of (a) hyperbolic secant-type and (b) Student's $t_3$-type , for $c=2.5$ and $T=10,1,0.01$.}\label{fig:other-psi}
\end{center}
\end{figure}

\begin{exam}[The Hyperbolic Secant Distribution]\mbox{\ \ \ }\\
The hyperbolic secant distribution is a symmetric distribution with exponential tails. The standardized density is equal to 
\begin{gather*}
h(r)=\frac{1}{2\cosh(r\pi/2)}.
\end{gather*}
The $\psi$-function of the corresponding (HS-type) redescending M-estimator is shown in Fig.~\ref{fig:other-psi}(a), for three different temperatures $(T=10,1,0.01)$. It is easy to show that $h(r)$ is rapidly varying at infinity. According to Proposition~\ref{prop:2}, the weight function $w_h(r;c,T)$ converges to $H(c-r)$  for $T\gegen 0$, and the corresponding M-estimator approaches the skipped mean.\hfill$\Box$
\end{exam}

\begin{exam}[Student's $t$-Distribution]\mbox{\ \ \ }\\ 
Student's $t$-distribution is a symmetric distribution with tails falling off according to a power law. The standardized density with $\nu>2$ degrees of freedom is equal to 
\begin{gather*}
t_\nu(r)=\frac{\Gamma((\nu+1)/2)}{\sqrt{\pi(\nu-2)}\,\Gamma(\nu/2)}\left(1+\frac{r^2}{\nu-2}\right)^{-(\nu+1)/2}.
\end{gather*}
The $\psi$-function of the corresponding ($t_\nu$-type) redescending M-estimator with $\nu=3$ is shown in Fig.~\ref{fig:other-psi}(b), for three different temperatures $(T=10,1,0.01)$.  The density $t_\nu(r)$ is regularly varying at infinity with index $\xi=-(\nu+1)$. From Proposition~2 follows:
\begin{gather*}
\lim_{T\gegen 0} w_{t_\nu}(r;c,T)=\frac{c^{\nu+1}}{c^{\nu+1}+r^{\nu+1}}.
\end{gather*}
For $\nu\gegen\infty$, this function approaches the step function $H(c-r)$.\hfill$\Box$
\end{exam}

\section{N-type M-estimators of location}\label{sec:ntype}

In this section the properties of the annealing M-estimator with a normal data model are explored. The effect of deterministic annealing on the objective function of the estimator is illustrated on a simple example with two clusters (data and outliers).

\subsection{Basic properties}\label{sec:basic}

\begin{figure}
\begin{center}
{\includescaled{0.94}{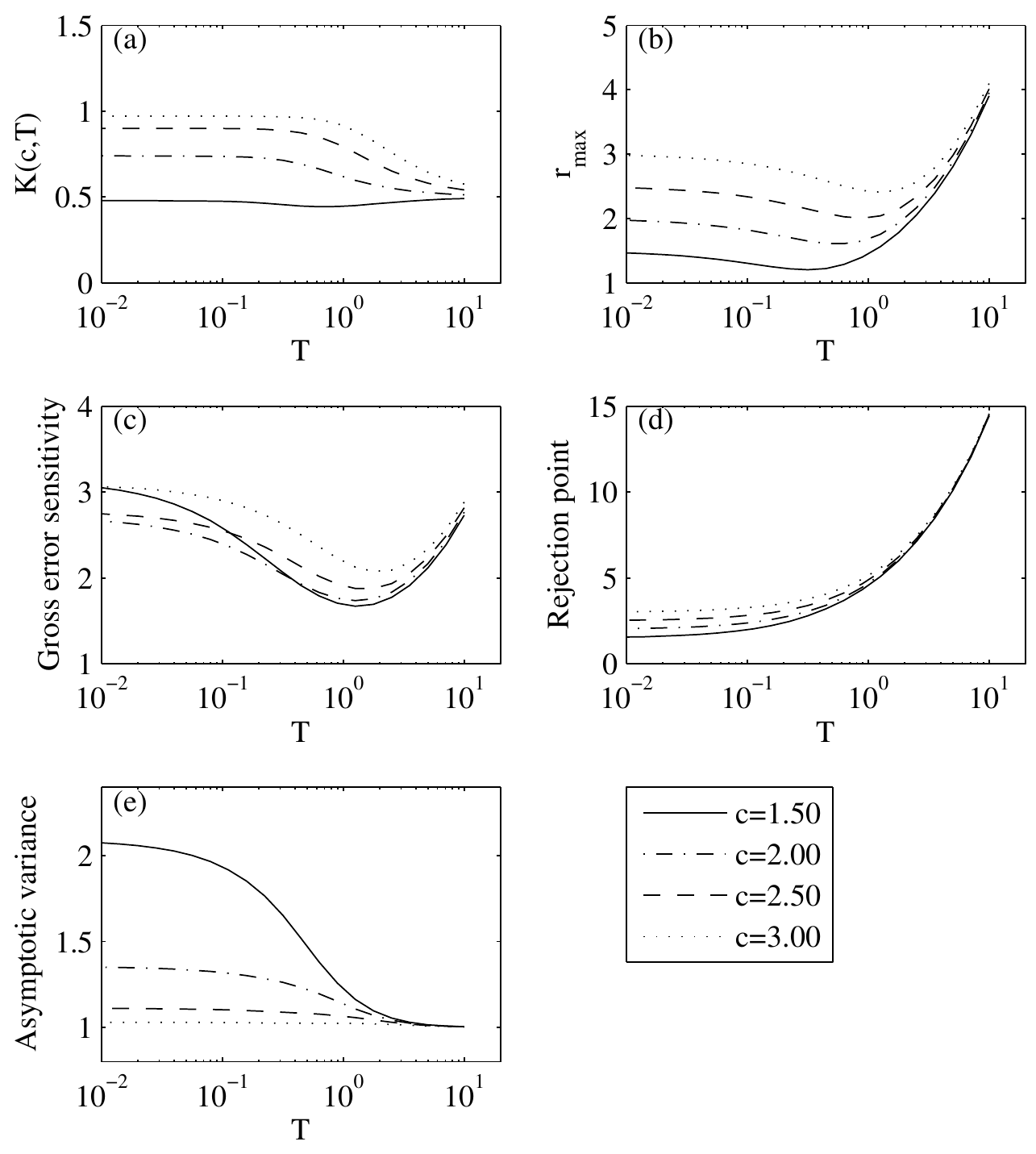}}
\caption{N-type M-estimator: (a) $K(c,T)$, (b) $\rmax$, (c) gross error sensitivity, (d) effective rejection point for $\varepsilon=10^{-3}$ and (e) asymptotic variance, as a function of the temperature $T$, for $c=1.5\!:\!0.5\!:\!3$.}
\label{fig:N-type-properties}
\end{center}
\end{figure}

The influence function is always proportional to $\psi$:
\begin{gather*}
\IF(r;\psi(r;c,T),F)=\psi(r;c,T)/K(c,T).
\end{gather*}
If the model distribution is the standard normal distribution, $K(c,T)$ is given by~\citep{Hampel}:
\begin{gather*}
K(c,T)=\int_{\R} r\,\psi(r;c,T)\,\phi(r)\dr = 2\int_{0}^\infty r\,\psi(r;c,T)\,\phi(r)\dr  .
\end{gather*}
Unfortunately, the integral cannot be written in closed form. Figure~\ref{fig:N-type-properties}(a) shows $K(c,T)$ as a function of $T$, for $c=1.5\!:\!0.5\!:\!3$, computed by numerical integration. Complications at very small values of $T$ can be avoided by splitting the interval of integration $[0,\infty)$ at $c$. The low- and high-temperature limits can be computed explicitly:
\begin{gather*}
\lim_{T\gegen 0} K(c,T)=2\,\Phi(c)-1-2\,c\,\phi(c),\quad \lim_{T\gegen\infty} K(c,T)=\frac{1}{2}.
\end{gather*}
The point of maximum influence can be computed by means of the Lambert $W$-function \citep{Corless}:
\begin{gather*}
\rmax(c,T)=\sqrt {2\,T\,\omega(c,T)+T},\quad\text{with}\quad\textstyle \omega(c,T)=W(\frac{1}{2}\exp(c^2/2T-1/2)).
\end{gather*}
$\rmax$ is shown in Figure~\ref{fig:N-type-properties}(b). The maximum value of the influence function is the gross-error sensitivity $\gamma^*$:
\begin{gather*}
\gamma^*(c,T)=\max_r\,\IF(r;c,T) = \frac{\psi(\rmax(c,T);c,T)}{K(c,T)}=\frac{1}{K(c,T)}
{\frac {2\,\sqrt {T}\,\omega(c,T)}{\sqrt {2\,\omega(c,T) +1}}}.
\end{gather*}
Figure~\ref{fig:N-type-properties}(c) shows the gross-error sensitivity as a function of $T$, for 
$c=1.5\!:\!0.5\!:\!3$. The minimum value lies in the range $1<T<2$, so if one aims to minimize $\gamma^*$, the final temperature should be chosen in that range. In the low-temperature limit we have
\begin{gather*}
\lim_{T\gegen 0} \gamma^*(c,T) =\frac{c}{2\,\Phi(c)-1-2\,c\,\phi(c)}.
\end{gather*}
At $T=0$, $\gamma^*$ is minimal for $c\approx 2.14$.
The weight function $w(r;c,T)$ is always positive, so the M-estimator does not have a finite rejection point. However, an effective rejection point can be computed for a threshold $\varepsilon$:
\begin{gather*}
\rp(c,T,\varepsilon)=\sup\, \{r:  \IF(r;c,T)>\varepsilon \}.
\end{gather*}
Figure~\ref{fig:N-type-properties}(d) shows the effective rejection point for $\varepsilon=10^{-3}$. In the limit $T\gegen 0$ the effective rejection point approaches the cutoff value $c$.
Finally, the asymptotic variance at the standard normal distribution, given by
\begin{gather*}
V(c,T)=\frac{\int_{\R} \psi(r;c,T)\,\phi(r)\dr}{K(c,T)^2},
\end{gather*}
is shown in Figure~\ref{fig:N-type-properties}(e). The low- and high-temperature limits are given by:
\begin{gather*}
\lim_{T\gegen 0} V(c,T)=\frac{1}{2\,\Phi(c)-1-2\,c\,\phi(c)},\quad \lim_{T\gegen\infty} V(c,T)=1.
\end{gather*}
Figure~\ref{fig:N-type-properties} shows that, for a given cutoff value $c$, it is not possible to minimize the gross-error sensitivity and the rejection point at the same time. The choice of the stopping temperature therefore depends on the problem at hand. If the asymptotic efficiency is important the cutoff value $c$ should be between 2.5 and 3, at the cost of a somewhat higher gross-error sensitivity and a larger rejection point. Cutoff values larger than 3 are not recommended.

\subsection{Effect of Deterministic Annealing}\label{sec:detann}

\begin{figure}[ht]
\begin{center}
{\includegraphics{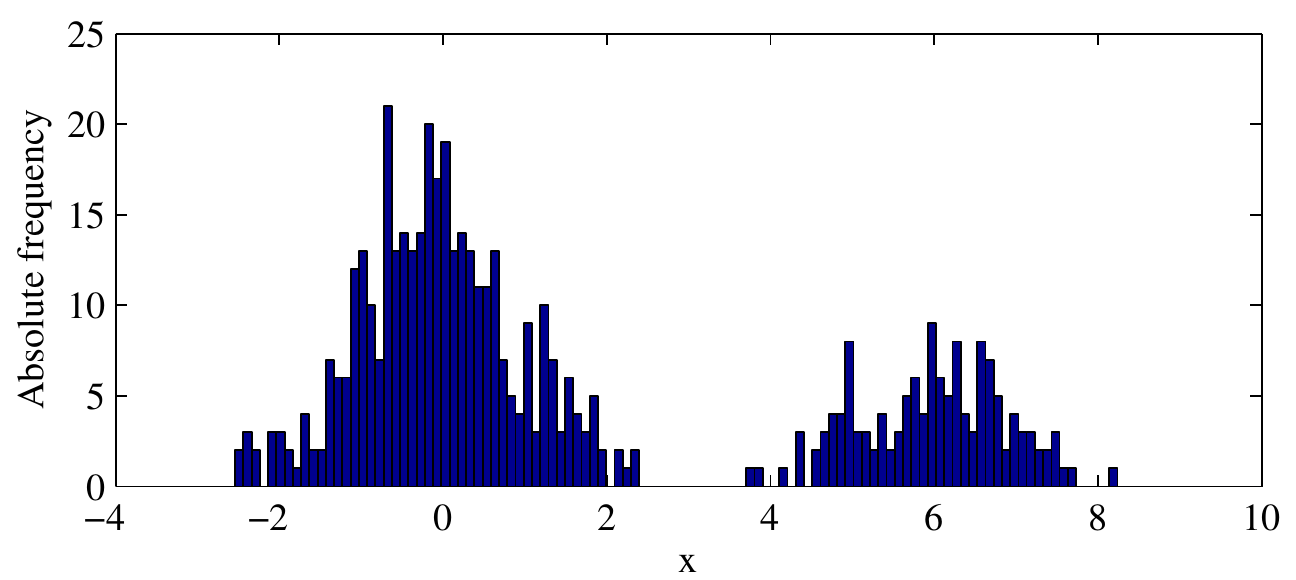}}
\caption{Example data set with 500 observations from a mixture of two standard normal distributions. The difference of the means is equal to six. There are 364 inliers and 136 outliers.}
\label{fig:sample}
\end{center}
\end{figure}

The effect of deterministic annealing on the minimization of the objective function of the N-type estimator is illustrated on a simple problem of location estimation with synthetic data. The data are generated from the following mixture model with mean-shift outliers (see Eq.~\eqref{eq:model}):
\begin{gather*}
h(x)=p\cdot\phi(x)+(1-p)\cdot\phi((x-m)/\sigma).
\end{gather*}
We have chosen the following mixture parameters:
\begin{gather*}
p=0.7, m=6, \sigma=1,
\end{gather*}
which results in two barely separated standard normal components. An example data set with 500 observations is shown in Figure~\ref{fig:sample}. There are 364 inliers and 136 outliers. 
The scale estimate $s$ is computed by taking the median of the absolute deviations from the half-sample mode, which in this situation is a better measure of the inlier location than the sample median~\citep{Bickel}. Its normal-consistent value for the example data is~1.31, whereas the normal-consistent MAD is equal to~1.56. The cutoff has been set to $c=2.5$.

\begin{figure}
\begin{center}
{\includescaled{1}{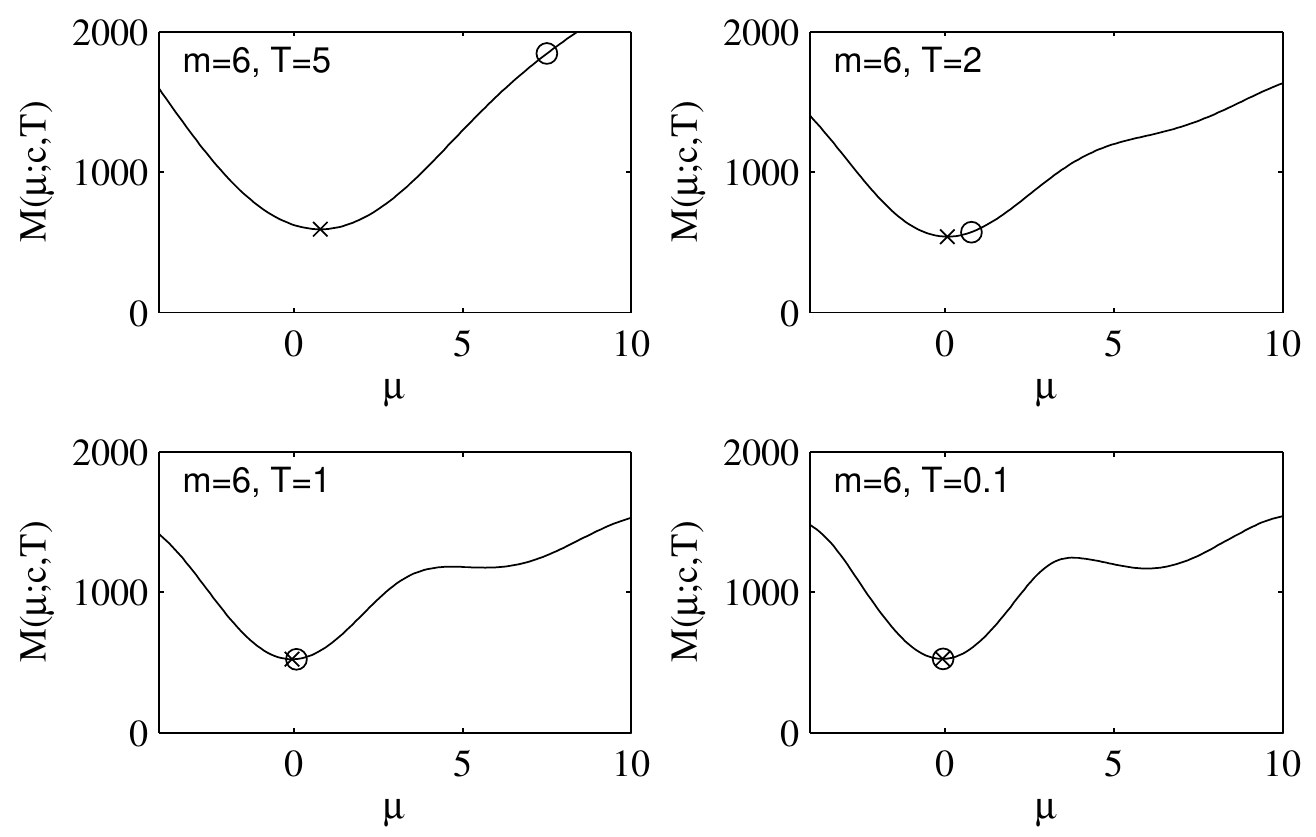}}
\caption{Evolution of the objective function $M(\mu;c,T)$ with falling temperature. The open circle (\usebox{\mycirc}) is the starting point of the iteration at the respective temperature, the x-mark (\usebox{\myxmark}) is the final estimate at the respective temperature.}
\label{fig:objfun}
\end{center}
\end{figure}

It is instructive to observe the evolution of the objective function 
$$M(\mu;c,T)=\sum_{i=1}^n \rho((x_i-\mu)/s;c,T)$$ 
with falling temperature $T$ (see Figure~\ref{fig:objfun}). At large $T$, the weights are nearly independent of the residuals, and the objective function is almost quadratic.
If the temperature is decreased, the objective function starts to reflect the structure of the data, eventually showing two clear local minima. These minima could be used to detect clusters in the data~\citep{GarlippMuller2003}. As the objective function is minimized at each temperature, the final estimate is now totally independent of the starting value. As long as the high-temperature minimum is closer to the deeper low-temperature minimum convergence to the latter is virtually guaranteed. 

\begin{figure}
\begin{center}
{\includescaled{1}{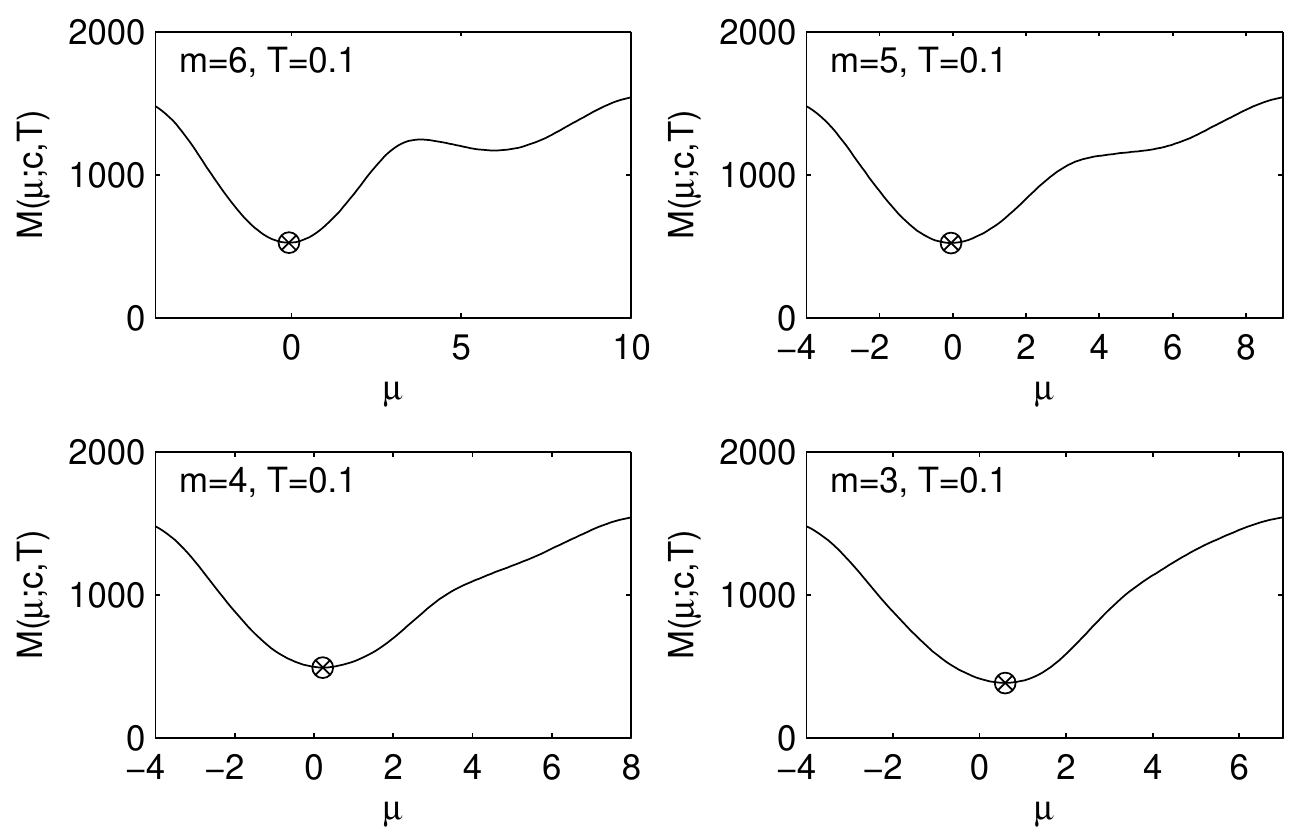}}
\caption{The objective function $M(\mu;c,T)$ at the final temperature $T=0.1$, for different values of the mean shift $m$ between inliers and outliers. The open circle (\usebox{\mycirc}) is the starting point of the iteration, the x-mark (\usebox{\myxmark}) is the final estimate.}
\label{fig:objfun_all}
\end{center}
\end{figure}

If the separation $m$ between inliers and outliers is decreased, the final objective function eventually has a single minimum. Figure~\ref{fig:objfun_all} shows the final objective function at $T=0.1$ for $m=6,5,4,3$. 
At $m=5$ the second local minimum has disappeared, but the objective function still has a point of inflection close to the outlier location, and the estimate is unbiased. At $m=4$ the point of inflection is barely visible, and the estimate shows a small bias. At $m=3$ the point of inflection has disappeared, and the estimate shows a clear bias. In contrast, the median and the half-sample mode are totally unaffected by the change in separation.

Deterministic annealing in combination with redescending M-estimators has already been proposed by~\citet{Li1996}. One of the weights function used there is a modified Welsch estimator with the weight function
\begin{gather*}
w(r;T)=\exp(-r^2/2T),
\end{gather*}
which is equal to the numerator of the N-type weight function in Eq.~\eqref{eq:w}. It is easy to show that the asymptotic variance of this estimator at the standard normal distribution is equal to
\begin{gather*}
V(T)=\frac{(1+T)^3}{(2+T)^{3/2}\,T^{3/2}},
\end{gather*}
and consequently
\begin{gather*}
\lim_{T\gegen 0} V(T)=\infty.
\end{gather*}
The same holds for the other two weight functions proposed by~\citet{Li1996}.

\section{Applications}\label{sec:applications}

In this section we present two application of the annealing M-estimator. In the first one the estimator is applied to the problem of estimating robustly the interaction vertex of a particle collision or a particle decay. The results show that annealing is instrumental in identifying and suppressing the outliers. In the second application the annealing M-estimator is used for regression diagnostics in the context of the estimation of the tail index of a distribution from a sample. 

\subsection{Robust regression and outlier detection}

The N-type estimator can be applied to robust regression with minimal modifications.  
The procedure is illustrated with the following problem from experimental particle physics. 

\begin{figure}
\begin{center}
\includescaled{0.2}{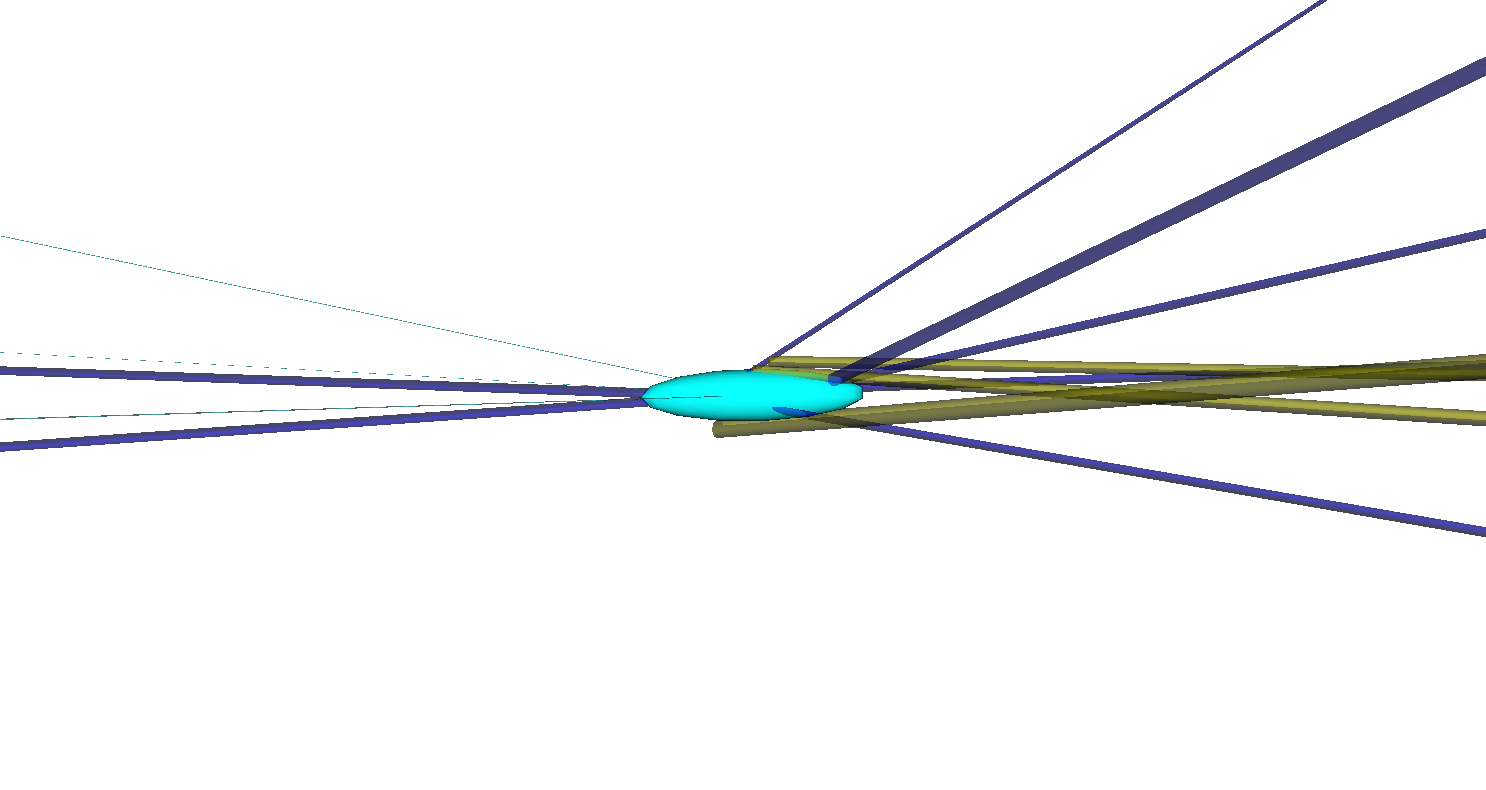}
\caption{A primary vertex with several outgoing tracks.}\label{fig:primary}
\end{center}
\end{figure}

An interaction vertex or briefly vertex is the point where particles are created by a collision of two other particles, or where an unstable particle decays and produces two or more daughter particles.
The position of the vertex has to be estimated from the parameters of the outgoing particles, the so-called track parameters. The track parameters consist of location, direction and curvature. They have to be estimated before the vertex can be estimated. As an illustration, Figure~\ref{fig:primary} shows a primary vertex, the interaction point of two beam particles in the accelerator, and several outgoing tracks. The precision of the estimated track parameters 
is indicated by the width of the tracks.

The least-squares (LS-)estimator of the vertex position $\vx$ minimizes the sum of the squared standardized
distances of all tracks from the vertex position $\vx$:
\begin{gather*}
\vhat_\mathrm{LS} = 
\argminv L(\vx) \with L(\vx)=\half\sum_{i=1}^{n} {r_i}^2(\vx)=\half\sum_{i=1}^{n} {{d_i}^2(\vx)}/{{\sigma_i}^2}.
\end{gather*}
The distance $d_i$ is approximated by an affine function of $\vx$, obtained by a first-order Taylor expansion of the track model, which is the solution of the equation of motion of the particle:
\begin{gather*}
d_i(\vx)\approx{}c_i+{\bm{a}_i}^T\vx.
\end{gather*}
The ${\sigma_i}^2$ are known from the estimation procedure of the track parameters.

With the redescending N-type M-estimator each track gets a weight $w_i$:
\begin{gather*}
w_i  = \frac{\exp(-{r_i}^2 /2T)} {\exp(-{r_i}^2/2T) + \exp(-\rcut^2/2T) }.
\end{gather*}
As a consequence, outlying or mis-measured tracks are downweighted by a factor $w_i$. As the factor $w_i$ depends on the current vertex position $\vx$, the M-estimator is computed as an iterated reweighted least-squares estimator. The dependence on the starting point is cured by annealing. The final weights can be used for a posterior classification of the tracks as inliers ($w_i>0.5$) or outliers ($w_i<0.5$). 

In our example we have used simulated events from the CMS experiment~\citep{fru:CMS94,CMS} at CERN, real data not yet being available. We have studied the estimation of the primary (beam-beam collision) vertex. For more details about the estimation problem, see~\citet{AVF}. The primary particles produced in the beam-beam collision are the inliers, whereas short-lived secondary particles produced in decays of unstable particles are the outliers, along with mis-measured primary tracks. Primary and secondary tracks can be identified from the simulation truth. Estimation of the primary vertex was done by the N-type M-estimator, with the least-squares estimator as the starting value. The annealing schedule was $T_0=256$, $T_{i+1}=\Tend+q (T_{i}-\Tend)$, with $q=0.25$.

The results are summarized in Table~\ref{tab:results}. The first column shows the type of annealing used, the second and third columns show the classification of the primary tracks by their final weights, the fourth and fifth columns show the classification of the secondary tracks, and the last column shows how many vertex estimates were within $100\,\mu{}m$ of the true vertex position, known from the simulation. Without annealing, the N-type M-estimator performs better at $T=1$ than at $T=0.01$. However, the results show that annealing is essential for the correct classification of primary and secondary tracks. The natural stopping temperature of the annealing procedure is $\Tend=1$, but cooling to $\Tend=0.01$ gives a slight improvement.

\begin{table}[b!]
\caption{Results of vertex estimation with the N-type M-estimator, using simulated data. For details see the text.}
\label{tab:results}
\begin{center}
\begin{tabular}{l|c|c|c|c|c}
       &\multicolumn{2}{|c}{inliers} & \multicolumn{2}{|c|}{outliers} & vertices\\ \hline 
Annealing schema & $w\!<\!0.5$ & $w\!>\!0.5$ & $w\!<\!0.5$ & $w\!>\!0.5$ & $n_\text{rec}$ \\ \hline 
  No annealing, $T=1$ & $0.312$ & $0.688$ & $0.859$ & $0.141$ & $1422$\\ 
  No annealing, $T=0.01$ & $0.512$ & $0.488$ & $0.899$ & $0.101$ & $1004$\\ 
  Annealing, $\Tend=1$ & $0.101$ & $0.899$ & $0.828$ & $0.172$ & $1913$\\ 
  Annealing, $\Tend=0.01$ & $0.092$ & $0.908$ & $0.829$ & $0.171$ & $1939$\\ 
\end{tabular}
\end{center}
\end{table}

A similar method can be employed for the estimation of the track parameters. In this case, several observations may compete for inclusion into the track, and the computation of the weights has to be modified accordingly~\citep{Fruh99}.

\subsection{Tail index estimation}

\begin{figure}
\begin{center}
\includegraphics{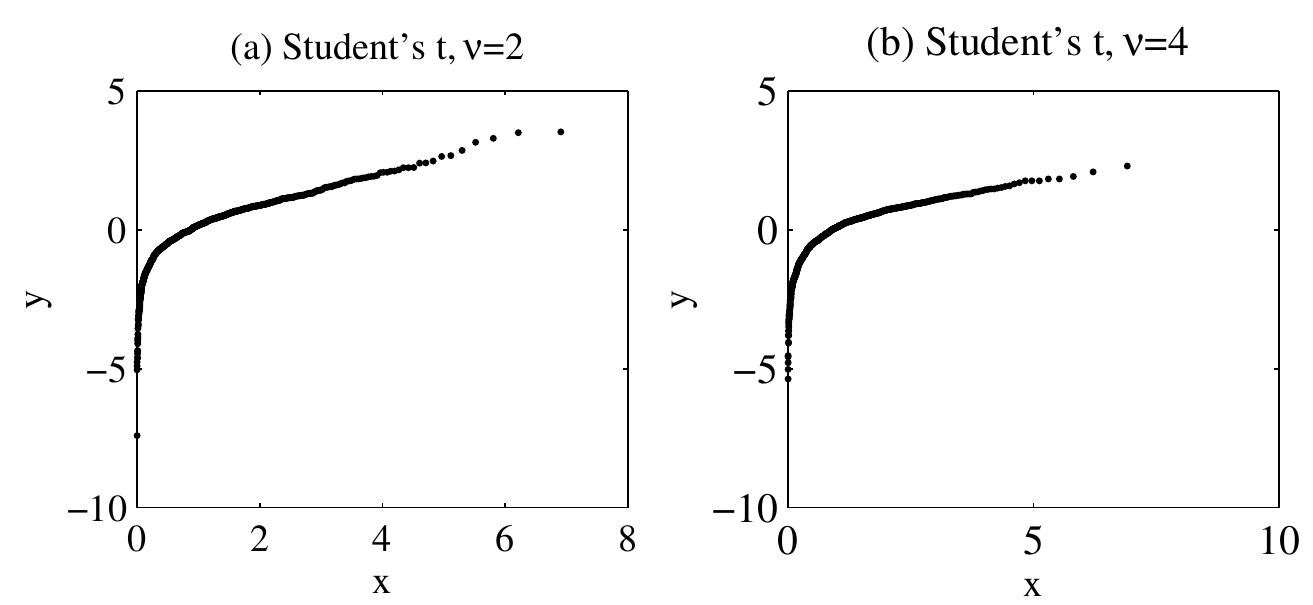}
\caption{Pareto quantile plots of two samples of size $n=1000$ from the $t$-distribution, with (a) $\nu=2$ and (b) $\nu=4$, respectively.}
\label{fig:pareto}
\end{center}
\end{figure}

The tail index $\alpha$ of the distribution of a random variable $X$ is defined by
\begin{gather*}
\alpha=\sup\{\delta>0:\mathrm{E}(|X|^\delta)<\infty\}.
\end{gather*}
The tail index determines how many moments of $X$ exist. A consistent estimator of $\alpha^{-1}$ from a sample $\xs$ is the Hill estimator~\citep{Hill}:
\begin{gather*}
{\hat{\alpha}_k}^{-1}=\frac{1}{k} \sum_{j=1}^k  \log X_{(n-j+1)}-\log X_{(n-k)}.
\end{gather*}
As pointed out by~\citet{Beirlant}, the choice of $k$ is a problem of regression diagnostics.
This can be understood by looking at the Pareto quantile plot of the sample. The latter is a
scatter plot $(x_j,y_j), j=1,\ldots,n$, with
\begin{gather*}
x_j=-\log\left(\frac{j}{n+1}\right),\quad y_j= \log X_{(n-j+1)},\quad j=1,\ldots,n.
\end{gather*}
As an example, Figure~\ref{fig:pareto} shows the Pareto quantile plot of two samples from the $t$-distribution, with $\nu=2$ and $\nu=4$, respectively. The sample size is $n=1000$.

If a line is fitted to the linear part of the plot, its slope is an estimate of $1/\alpha$.
The problem is therefore to find the linear part of the plot. It is worth noting that standard robust regression methods such as LMS or LTS~\citep{Rousseeuw} will fail, as by definition the tail is not the majority of the data.

The N-type M-estimator can be used for regression diagnostics in order to find the linear part of the Pareto quantile plot. The algorithm is based on the idea of the forward search~\citep{AtkinsonRiani} and is composed of the following steps:
\paragraph{Algorithm A}
\begin{enumerate}
\item[A1.] Compute the scale of $y_i$, using the asymptotic expression for quantiles and a kernel estimator for the probability density function. As the kernel estimator is unreliable in the very extreme part of the tail, the largest half percent of the sample is discarded.
\item[A2.] Fit a robust regression line with the N-type M-estimator to the $m$ largest points in the Pareto quantile plot. The starting line is the LMS regression line. The temperature is set to $T=1$.
\item[A3.] Freeze all weights and extend the fit successively to the lower portion of the plot, by adding $m$ points at a time. The choice of $m$ is a trade-off between speed and safety.
\item[A4.] Stop adding points when the new weights get too small, indicating failure of the linear model.
\end{enumerate}

\begin{figure}
\begin{center}
\includegraphics{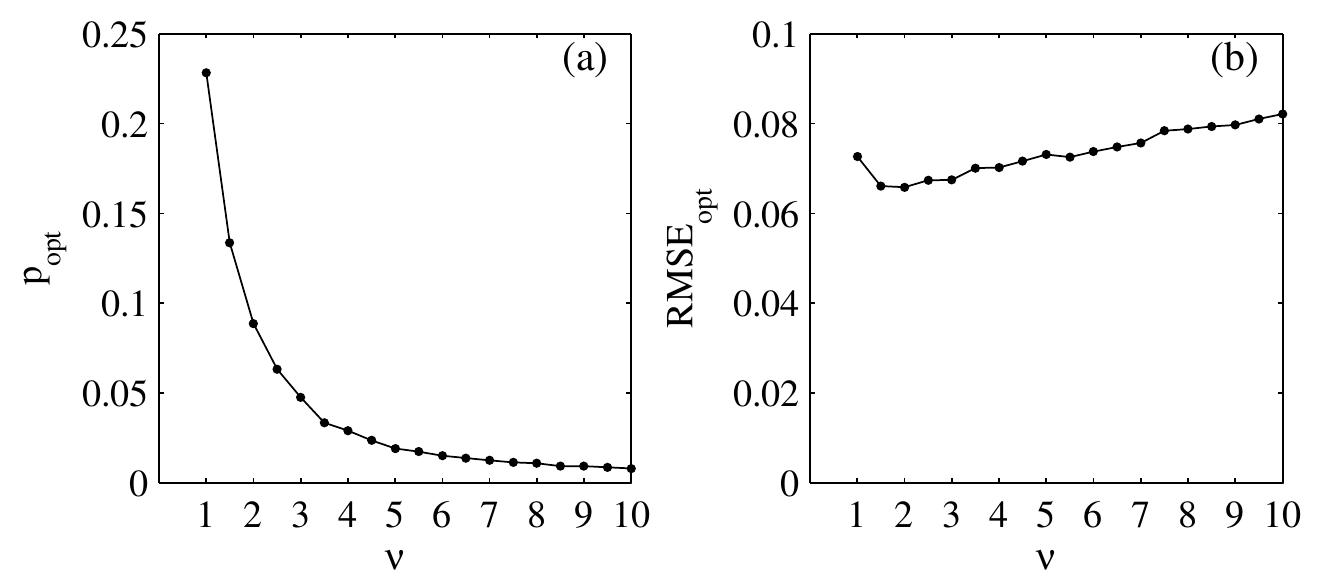}
\caption{(a) Optimal proportion $p_{\mathrm{opt}}$ of the sample and (b) optimal RMSE of the Hill estimator, as a function of $\nu$.}
\label{fig:tail_estimator_hill}
\end{center}
\end{figure}

We have tested the algorithm on samples from the $t$-distribution with $\nu$ degrees of freedom, with $n=1000$ and 
$\nu=1\!:\!0.5\!:\!10$. The baseline is the Hill estimator using the optimal value of $k$. The latter was found for each value of $\nu$ by computing Hill estimators with different values of $k$ and choosing the one that minimizes the root mean-square error (RMSE) of ${\hat{\alpha}_k}^{-1}$ with respect to the true value ${\alpha}^{-1}=1/\nu$. Figure~\ref{fig:tail_estimator_hill} shows the optimal proportion $p=k/n$ and the RMSE of the corresponding Hill estimators as a function of~$\nu$.

\begin{figure}
\begin{center}
\includegraphics{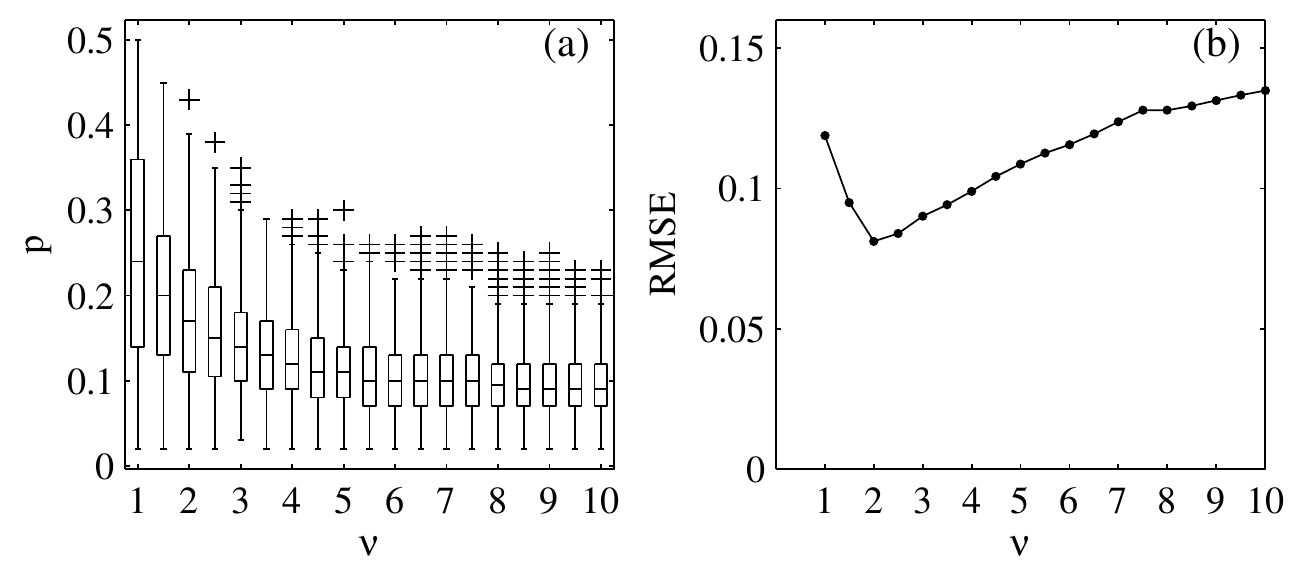}
\caption{(a) Proportion $p$ of the sample used by Algorithm~A and (b) resulting RMSE of Algorithm~A, as a function of $\nu$.}
\label{fig:tail_estimator_ada}
\end{center}
\end{figure}

Algorithm A as described above was run with $m=10$, i.e.~one percent of the sample size. The cutoff parameter $c$ was adjusted at the 99\%-quantile  of the $\chi^2_1$ distribution, i.e.~at $c=2.576$. The fit was stopped as soon as at least half of the new weights were smaller than 99\% of the maximum weight $w_{\mathrm{max}}=1/[1+\exp(-c^2/2)]=0.965$.
Figure~\ref{fig:tail_estimator_ada} summarizes the results. The left hand panel (a) shows box plots of the proportion of the sample included in the regression, one for each value of $\nu$. The right hand panel (b) shows the resulting RMSE of ${\hat{\alpha}_k}^{-1}$ with respect to the true value ${\alpha}^{-1}=1/\nu$, for all $\nu$. The figure clearly shows that there is a tendency to include more data points than required for the optimal estimate. As a consequence, the RMSE is somewhat larger than in the optimal case. On the other hand, in a real life situation no external information at all may be available about the optimal value of $k$. In this case the regression diagnostics approach, which is entirely driven by the data, is a viable alternative.

\section{Summary}

A new type of redescending M-estimators has been introduced, suitable for combination with deterministic annealing.
It has been shown that the annealing M-estimator converges to the skipped mean if and only if the inlier density is rapidly varying at infinity. Deterministic annealing helps to make the estimator insensitive to the starting point of iteration. Possible applications are location estimation, robust regression and regression diagnostics. The new type of estimators is particularly useful if the scale of the observations is known. In other cases the scale has to be estimated from the data, preferably in a robust way. 


\begin{thebibliography}{17}

\bibitem[{A}tkinson and {R}iani(2000)]{AtkinsonRiani}
{A}tkinson, A., and {R}iani, M. (2000).
\newblock \emph{Robust Diagnostic Regression Analysis}.
\newblock Springer, New York.

\bibitem[{B}eirlant et~al.(1996){B}eirlant, {V}ynckier and
  {T}eugels]{Beirlant}
{B}eirlant, J., {V}ynckier, P., and {T}eugels, J.~L. (1996).
\newblock Tail index estimation, Pareto quantile plots, and regression
  diagnostics.
\newblock \emph{Journal of the American Statistical Asssociation}, 91\penalty0
  (436):1659.

\bibitem[{B}ickel and {F}{r\"uhwirth}(2006)]{Bickel}
{B}ickel, D.~R., and {F}{r\"uhwirth}, R. (2006).
\newblock On a fast, robust estimator of the mode: Comparisons to other robust
  estimators with applications.
\newblock \emph{Computational Statistics and Data Analysis}, 50:3500.

\bibitem[\protect\citeauthoryear{{CMS Collaboration}}{{CMS
  Collaboration}}{1994}]{fru:CMS94}
CMS collaboration (1994).
\newblock \emph{CMS Technical proposal}.
\newblock Technical Report CERN/LHCC 94-38, CERN, Geneva.

\bibitem[\protect\citeauthoryear{{CMS Collaboration}}{{CMS
  Collaboration}}{2007}]{CMS}
{CMS Collaboration} (2007).
\newblock {CMS Detector Information}. URL:\\
  {\tt http://cmsinfo.cern.ch/outreach/CMSdetectorInfo/CMSdetectorInfo.html}.

\bibitem[{C}orless~et al.(1996)]{Corless}
{C}orless, R.~M., ~et al. (1996).
\newblock On the {L}ambert {W} {F}unction (1996).
\newblock \emph{Advances in Computational Mathematics}, 5:329.

\bibitem[Dempster et~al.(1977)Dempster, Laird, and Rubin]{emalgo}
Dempster, A.~P., Laird, N.~M., and Rubin, D.~B. (1977).
\newblock {M}aximum likelihood from incomplete data via the {EM} algorithm.
\newblock \emph{Journal of the Royal Statistical Society B}, 39:1.

\bibitem[{F}{r\"uhwirth} and {S}trandlie(1999)]{Fruh99}
{F}{r\"uhwirth}, R., and {S}trandlie, A. (1999).
\newblock Track fitting with ambiguities and noise: a study of elastic tracking and nonlinear filters.
\newblock \emph{Computer Physics Communications}, 120:197.

\bibitem[{G}arlipp and {M}{\"u}ller(2005)]{GarlippMuller2003}
{G}arlipp, T., and {M}{\"u}ller, Ch. (2005).
\newblock Regression clustering with redescending {M}-estimators.
\newblock In D.~{B}aier and K.-D. {W}ernecke, editors, \emph{Innovations in
  Classification, Data Science, and Information Systems}. Springer, Berlin, Heidelberg,
  New York. 

\bibitem[{H}ampel~et al.(1986)]{Hampel}
{H}ampel, F.~R., ~et al. (1986).
\newblock \emph{{R}obust {S}tatistics: {T}he {A}pproach {B}ased on {I}nfluence
  {F}unctions}.
\newblock John Wiley \& Sons, New York.

\bibitem[{H}ill(1975)]{Hill}
{H}ill, B.~M. (1975).
\newblock A simple general approach to inference about the tail of a
  distribution.
\newblock \emph{The Annals of Statistics}, 3(5):1163.

\bibitem[Huber(2004)]{Huber}
Huber, P.~J. (2004).
\newblock \emph{{R}obust {S}tatistics: {T}heory and {M}ethods}.
\newblock John Wiley \& Sons, New York.

\bibitem[{L}i(1996)]{Li1996}
{L}i, S.~Z. (1996).
\newblock Robustizing robust {M}-estimation using deterministic annealing.
\newblock \emph{Pattern recognition}, 29(1):159.

\bibitem[{M}{\"u}ller(2004)]{Muller2004}
{M}{\"u}ller, Ch. (2004).
\newblock Redescending {M}-estimators in regression analysis, cluster analysis
  and image analysis.
\newblock \emph{Discussiones Mathematicae --- Probability and Statistics},
  24:59.


\bibitem[{R}ose(1998)]{Rose1998}
{R}ose, K. (1998).
\newblock Deterministic annealing for clustering, compression, classification,
  regression, and related optimization problems.
\newblock \emph{Proceedings of the IEEE}, 86(11):2210.

\bibitem[Rousseeuw and Leroy(1987)]{Rousseeuw}
Rousseeuw, P.~J., and Leroy, A.~M. (1987).
\newblock \emph{{R}obust {R}egression and {O}utlier {D}etection}.
\newblock John Wiley \& Sons, New York.

\bibitem[Seneta(1976)]{Seneta}
Seneta, E. (1976).
\newblock \emph{{R}egularly {V}arying {F}unctions}.
\newblock Springer, Berlin, Heidelberg, New York.

\bibitem[{W}altenberger et~al.(2007){W}altenberger, {F}r{\"u}hwirth and
  {V}anlaer]{AVF}
{W}altenberger, W., {F}r{\"u}hwirth, R., and {V}anlaer, P. (2007).
\newblock Adaptive vertex fitting.
\newblock \emph{Journal of Physics G: Nuclear and Particle Physics},
  34:N343.

\end{thebibliography}

\end{document}